\documentclass{article}
\usepackage[preprint]{colm2026_conference}

\usepackage{microtype}
\usepackage{hyperref}
\usepackage{url}
\usepackage{booktabs}
\usepackage{amsmath}
\usepackage{multirow}
\usepackage{makecell}
\usepackage{tabularx}

\usepackage[utf8]{inputenc}
\usepackage{longtable}

\usepackage{lineno}

\definecolor{darkblue}{rgb}{0, 0, 0.5}
\hypersetup{colorlinks=true, citecolor=darkblue, linkcolor=darkblue, urlcolor=darkblue}

 \title{The \textit{Silicon Society} Cookbook:\\
 Design Space of LLM-based Social Simulations}

\author{
    \begin{tabular}{@{}ccc@{}}
        Aurélien Bück-Kaeffer$^{1,2,4}$ & Sneheel Sarangi$^{1,2}$ & Maximilian Puelma Touzel$^{1,3}$ \\
        Reihaneh Rabbany$^{1,2}$ & Zachary Yang$^{1,2,4}$ & Jean-François Godbout$^{2,3}$ \\
    \end{tabular}\\[6pt]
    $^1$ McGill University, $^2$ Mila - Quebec Artificial Intelligence Institute, \\
    $^3$ Université de Montréal, $^4$ Ubisoft La Forge
}

\begin{document}

\maketitle

\begin{abstract}
Studies attempting to simulate human behavior with \textit{Silicon Societies} grow in numbers while LLM-only social networks have started appearing outside of controlled settings. However, the design space of these networks remains under-studied, which contributes to a gap in validating model realism. To enable future works to make more informed design decisions, we perform a systematic analysis of the consequences and interactions of key design choices in simulated social networks, including the choice of base model used to model individual agents, and how they are connected to each other. Using surveys as a proxy for agent opinions, our findings suggest that the geometry of the design space is non-trivial, with some parameters behaving in additive ways while others display more complex interactions. In particular, the choice of the base LLM is the most important variable impacting the simulation outcomes.
\end{abstract}

\section{Introduction}

Agent-Based Models (ABMs) and social simulations have a long history in social sciences \citep{de2014agent}, but recent years have seen the emergence of a new paradigm using Large Language Models to simulate human interactions. The promise of models encompassing both micro- (individual users, messages, or even activations for a given token) and macro-level phenomena (echo chambers, herd effects, information propagation) at arbitrary levels of precision and scale has proven alluring to many (See Section \ref{related_works} for a survey). However, while new increasingly more complex simulators are continuously being published, a chronic lack of validation characterizing many recent works limits their ability to build upon each other and address the limitations of classical ABMs \citep{li2025llmgeneratedpersonapromise,larooij2025largelanguagemodelssolve}.

In this paper, we are specifically interested in \textit{Silicon Societies}, to which we limit the scope of our work. We define a \textit{Silicon Society} as a simulation of human interactions using Large Language Models. This definition includes controlled studies seeking to reproduce some traits of human social interaction networks (e.g. OASIS, \cite{yang2025oasisopenagentsocial}), but to a lesser extent also covers naturally emerging social networks populated by LLMs (e.g. Moltbook). We explicitly exclude multi-agent frameworks seeking to achieve a particular goal (e.g. coding agents) other than human-likeness, even though we do note that some of the validation techniques they developed may be transferable to this setting.

Validating \textit{Silicon Societies} is a  non-trivial task. How does one go about defining and then measuring human-likeness on the scale of a social network? Numerous attempts have been made, with ``varying levels of persuasiveness and rigor'' \citep{larooij2025largelanguagemodelssolve}. As of now, a standard has yet to emerge. Human baselines, which are often application-specific, are challenging to obtain. 
Instead of attempting to build the "most realistic" simulator according to arbitrary metrics, in this paper  we seek to understand how the parameters we pick when designing a \textit{Silicon Society} impacts the simulation and its outcome.
Mapping the design space of these social simulations enables future work to make more informed design decisions. 
To this end, we perform a systematic analysis of simulation trajectories over 595 roll-outs with some parameters randomized to identify the nature and intensity of their respective impacts, and their complex interactions.

We hypothesize that the design space will not behave additively (given that parameter A and parameter B both independently increase metric X, the space behaves additively if a simulation with both parameters A and B has a larger metric X than either of them separately). Instead, we expect to find that parameters will display complex interactions, making it difficult to predict how a simulation will behave based solely on its initial settings. We further hypothesize that fine-tuning LLMs on social media data will make them stylistically closer to humans and change their behavior in simulations.

We conduct a systematic study of 595 simulation roll-outs across a Silicon Society design space spanning seven parameters. We use surveyed opinions as an operational proxy for agent beliefs. Our findings reveal that the design space is neither uniformly additive nor uniformly complex. The choice of base model is the single most influential factor, dominating both stylistic indistinguishability ($\eta^2$=0.266 on BERT AI-detectability) and opinion dynamics ($\eta^2$=0.090 on net consensus change). Fine-tuning on social media data substantially correlates with stronger social dynamics and reduces AI detectability, while a biased news agent has no measurable effect on any metric. Several interactions prove model-dependent and scale-dependent, underscoring that simulation behavior cannot be reliably predicted from marginal effects alone.

Our main contributions are \textbf{1)} measuring the relative importance of several \textit{Silicon Society} design decisions (allowing us to identify the most important ones) and the nature of their respective impacts on simulation roll-outs, \textbf{2)} characterizing the partially-additive nature of the design space, and \textbf{3)} providing evidence that fine-tuning models on social media data leads to simulations stylistically closer to human social media text and intensified opinion dynamics during simulations.

\section{Related works}
\label{related_works}

\textbf{LLM-based social simulators} have gained a significant amount of traction, and works in this field are
numerous.
\cite{larooij2025largelanguagemodelssolve} survey at least 35 such simulators with human imitation objectives. Notable works include OASIS \citep{yang2025oasisopenagentsocial} which achieves 1M agents, performs ablation studies over some of the simulation components, and observes convincing macro-scale effects (echo chambers, herding effect). Google Deepmind's Concordia \citep{vezhnevets2023generative} introduced modular cognitive agent components. Other notable recent works include \cite{zheng2026rolesimllm,lu2026real} and \cite{yuzhe2026twinmarket}, each following different simulation setting and validation. 
 The social simulation field is growing too fast 
 for validation and methodological standards to be met by wide adoption (on March 31, 2026, searching "LLM social simulation" in Google Scholar returned 4200 results for 2026 alone). 
In parallel, and while not strictly speaking "simulators", LLM-populated social networks have started appearing outside of controlled settings. Most notably, Moltbook has already been the subject of many articles  \citep{sodano2026emergencefragilityllmbasedsocial,jiang2026humans,holtz2026anatomy}.

\textbf{The validation gap} surrounding the aforementioned social simulators has been mentioned on occasion, yet remains heavily under-addressed in relation to the amount of articles being published. The recent survey by \cite{larooij2025largelanguagemodelssolve} discusses how poor standards prevent LLM-based social simulation from addressing long-standing classical Agent-Based models weaknesses. \cite{li2025llmgeneratedpersonapromise} outline the necessity of a science of persona creations. More recently, \cite{li2026positionaiagentsyet} shared their position regarding the mismatch between the way models are validated and what simulation-as-a-science requires. \cite{seshadri2026lost} claim that "current evaluation practices risk misrepresenting agent capabilities across diverse user populations and may obscure real-world deployment challenges".

Our study attempts to begin addressing the validation gap in LLM-based Social Simulators by building towards a better understanding of the design space of \textit{Silicon Societies}.
The goal is to better  understand how the decisions we make when building these complex systems affect the dynamics that occur within them. 

\section{Experiments}

\subsection{Simulator}

In order to avoid being constrained into some pre-existing simulator's assumptions, which might prevent us from conducting our study on all the parameters we are interested in, we build our own social network simulator to have full control over all assumptions.

Our design is loosely based on \cite{vezhnevets2023generative}. We instantiate a set of agents to populate our simulation and create a followership network connecting them together. The specific way in which this is done depends on the simulation parameters (see \ref{sim_parameters}). Agents get attributed an "action probability" fixed for the duration of the simulation. At each simulation step, 10 agents are selected according to this action probability. The first 9 observe a recent thread in which someone they follow participated, while the last one has a 1/3 probability of creating a new thread and a 2/3 probability of replying to an already existing one. For details on the simulator, see \ref{appendix_simulator_details}. The simulations run for a total of 2500 steps, with the agents being surveyed every 250 steps. We select 3 possible survey questions. In order to make opinion dynamics as clear as possible, we chose questions that are maximally divisive (see \ref{appendix_designing_survey_questions}). Using BluePrint \citep{bückkaeffer2025textttblueprintsocialmediauser} we find that there are on average 24.3 messages per user per month being posted, meaning simulating 2500 messages between 1024 agents corresponds to about 75h in simulated time. Each simulation runs on a A100 GPU and takes between 2 and 11h to complete depending on the specific parameters.

\subsection{Variables}
\label{sim_parameters}

Our study seeks to measure the relative impact and interactions of the seven following variables on the various simulation metrics:

\begin{itemize}
    \item The \textbf{number of agents}. The options are 64, 256, 1024 and 4096.
    \item The \textbf{base model} that we finetune using the BluePrint social media data to obtain the agents in our simulation (see \ref{finetuning_on_social_media}). The options are \textit{Llama-3.1-Minitaur-8B}, \textit{Llama-3.1-8B}, \textit{Qwen2.5-7B-Instruct} and \textit{gemma-3-4b-pt} \citep{binz2024centaurfoundationmodelhuman,qwen2,gemma_2025}.
    \item The \textbf{network topology}, i.e., how the follower graph is initialized. The options are an ER model \citep{erdos1959randomgraphs} or a directed scale-free graph \citep{10.5555/644108.644133}. \citet{li2024largelanguagemodeldrivenmultiagent} already do some similar analyses.
    \item Whether the network is initialized with \textbf{homophily} or not. If "Yes", agents with similar initial opinions will tend to be placed closer together in the network (\ref{appendix_homophily}).
    \item Whether models \textbf{know about their own survey answers} or not. If "Yes", survey questions and the model's answer are added to the model's context.
    \item Whether a biased \textbf{news agent} is present in the simulation or not. If "Yes", a news agent is placed at the node with the highest degree in the followership network. It creates "news" posts pre-generated using Gemini 3.1 explicitly biased in favor of one of the survey options.
    \item \textbf{Proportions}. Options are \textit{BluePrint}, \textit{Uniform}, \textit{Distribution} and \textit{Average}.
\end{itemize}

\textbf{Proportions} refers to the fact that for each base model, we finetuned 25 different LoRA adapters on social media data (one for each persona cluster, see \ref{finetuning_on_social_media}). Given that there are more than 25 agents in the simulations, each LoRA is used to run some portion of the population. If every LoRA is responsible for 1/25th of the population, we say that we are using \textit{Uniform} proportions. If instead we used the proportions from the BluePrint dataset (e.g. if cluster 3 made up 12\% of the BluePrint dataset, then 12\% of the agents in our simulation will be run using LoRA number 3), we say that we are using \textit{BluePrint} proportions. The next two options (\textit{Distribution} and \textit{Average}) attempt to study whether it is important that LLMs imitate the full distribution of human opinions, or if simply giving the most likely average human answer is enough. While previous works discuss this issue \citep{jiang2025artificialhivemindopenendedhomogeneity}, it has not, to our knowledge, been directly tested in social simulation environments before.

Consider the SimBench  benchmark \citep{hu2025simbenchbenchmarkingabilitylarge} which measures alignment of LLMs against human opinions by asking a multiple choice question, collecting a vector containing the probability that the model gives each possible answer, and comparing it against the distribution of human answers to that same question. We define a matrix $H$ that captures the human baseline opinions, where $H_{ij}$ is the probability that a human gives answer $j$ to question $i$ in SimBench. Similarly, matrix $M_\ell$ is the opinion matrix of model $\ell$, where $M_{\ell:ij}$ is the probability that model $\ell$ gives answer $j$ to question $i$. Aligning $\ell$ with human opinion distribution can then be expressed as minimizing some matrix norm $||H-M_\ell||$.

Now if we have a population $L$ of distinct models at our disposal, we may align that population's opinion distribution with the human distribution by optimizing a convex weighted sum.
$$\min_{w_\ell\forall\ell\in L}||H-\sum_{\ell\in L}w_\ell M_\ell|| \text{ s.t } \sum_{\ell\in L}w_\ell = 1, w_\ell\geq0\forall\ell\in L$$

Essentially, we are using a set of models as a vector basis to "span the space of opinions". Using the weights $w_\ell$ obtained this way in our simulator, we say that we are using \textit{Distribution} proportions, because we match human opinions in distribution. If instead we replace $H$ by $\hat{H}$, in which for each row of $H$ we set the maximum element to 1 and all the others to 0, and optimize for this new target, we say that we are using \textit{Average} proportions, since we match the average (most common) human opinions. Intuitively, we test whether matching the full distribution of human opinions matters more than matching only the most common answer. Notably, both \textit{Distribution} and \textit{Average} yield SimBench scores higher than any individual model in the population (for Minitaur, we obtain respectively 17.08 and 17.05, while the best individual model stands at 14.60) despite very different weight distributions.

In total, we performed 595 simulation roll-outs with randomized parameters among the options listed (\ref{appendix_sim_counts}). While it is clear that each of these variables could comprise many more options, and that many more variables could be considered, there are already 1024 possible settings combinations with our current set. We acknowledge that this is a limitation, and only attempt to narrow some parts of the design space, which future works can build upon.

\subsection{Metrics} \label{metrics}

Evaluating the validity of \textit{Silicon Societies} is a challenging endeavor. Due to the imitative nature of their objective, one has to define what being "realistic" means. The literature has yet to converge on a set of metrics, and \citet{larooij2025largelanguagemodelssolve} point out that "Validation remains poorly addressed, with many studies relying solely on subjective assessments of model ‘believability’, and even the most rigorous validation failing to adequately evidence operational validity". Our only "realism" metric consists in a BERT classifier (based on \cite{zhang2022twhin}) we trained to distinguish between real human-written message threads and LLM-generated simulation threads (using BluePrint data as human samples and threads from all of our simulations as LLM data, totaling 200000 training samples and carefully avoiding data leakage). For each simulation around 150 threads are use as a test set. If a thread is harder to identify as being AI-generated, we interpret it as a sign suggesting that it is more realistic.

All other metrics are computed using model survey responses. Considering that they lack human baselines, they \textbf{should not} be used to claim that a given set of setting is more realistic than some other. They can, however, be used to measure how each setting impacts the simulation's opinion dynamics, giving us clues regarding which parameters matter most and how they interact. The complete list of metrics is available in \ref{appendix_metrics}, we invite readers to refer to it as needed.

\section{Fine-tuning on Social Media Data}
\label{finetuning_on_social_media}

LLMs are infamous for their characteristic "AI" style of writing: sometimes robotic, or over-agreeable, their distinctive sentence structures and vocabulary makes them \textit{feel} artificial \citep{Reinhart_2025}. Instruction-tuned commercial models often actively avoid presenting personal opinions, especially on controversial topics. By design, LLMs are helpful, pleasant and non-problematic. Notably, these are not the traits which characterize social media discourse and interactions. In an attempt to correct these discrepancies, we use social media data from the BluePrint dataset \citep{bückkaeffer2025textttblueprintsocialmediauser} to finetune models using LoRAs \citep{hu2021loralowrankadaptationlarge}. While providing persona descriptions is a more common approach due to it's relative low cost and straightforwardness, some evidence suggests that demographic-based persona descriptions may make LLMs behave in stereotypical ways \citep{venkit2026needsociallygroundedpersonaframework} while more verbose descriptions could lead to unrealistic outcomes \citep{li2025llmgeneratedpersonapromise}, motivating our fine-tuning solution.

The BluePrint paper uses user histories to compute "user embeddings", then clusters similar users together into persona archetypes. Training LLMs on each of these archetypes presumably creates models covering a wide range of user behaviors. While we use this dataset, we make some changes to the approach of \cite{bückkaeffer2025textttblueprintsocialmediauser}. Instead of only using messages in order to compute each user embedding, we also include embeddings from posts the user positively interacted with (likes, reposts). This change allows us to consider more user behaviors into our clustering process and include users which do not write posts yet interact through other actions. We finetune four different models: \textit{Llama-3.1-Minitaur-8B} (a "foundation model for human cognition", \cite{binz2024centaurfoundationmodelhuman}), \textit{Llama-3.1-8B} (to isolate the impact of \cite{binz2024centaurfoundationmodelhuman}), \textit{gemma-3-4b-pt} and \textit{Qwen-2.5-7B-Instruct}. Among those, only Qwen is instruction tuned: that is because there is some limited evidence suggesting that instruction tuning may collapse answer diversity \citep{shypuladoes}, and we expect diversity to matter when simulating social interactions.

In order to measure the results of this training, we compare the models on SimBench \citep{hu2025simbenchbenchmarkingabilitylarge} before and after fine-tuning (Table \ref{tab:lora_scores_std}). Overall, it appears that models tend to obtain better scores after fine-tuning, suggesting that training models on social media data increases their alignment with human opinions. If the model already performed well, as for Minitaur, the effect is either negligible or slightly negative. While these results serve to justify our fine-tuning approach, it is important to keep in mind that benchmarks are inherently gameable, and only provide a proxy to the underlying properties of the models. Notably, our later results do not suggest that SimBench score correlate with more stylistic similarity with humans, or with any of the opinion dynamics metrics we measured in our experiments. For a stronger justification of our approach, we present an ablation study of this training in section \ref{blueprint_results}.

\begin{table}[h]
\centering
\begin{tabular}{lrrr}
\toprule
Base Model & Base SimBench Score & BluePrint LoRA score $\mu \pm \sigma$ & Avg. $\Delta$ \\
\midrule
Llama-3.1-Minitaur-8B & 13.23 & $12.46\pm1.56$ & $-0.77$ \\
gemma-3-4b-pt         & $-3.42$ & $1.78\pm4.72$  & $+5.20$ \\
Llama-3.1-8B          & $-0.06$ & $1.14\pm4.51$  & $+1.20$ \\
Qwen-2.5-7B-Instruct  & $-34.87$ & $-26.87\pm11.90$ & $+8.00$ \\
\bottomrule
\end{tabular}
\caption{Comparing SimBench scores with vs without LoRA fine-tuning on BluePrint. For each base model, we train 25 loras and take the average and standard deviation of their scores. Fine-tuning on social media data using BluePrint gives models better SimBench scores on average, unless they were strong already (Minitaur) in which case the effect is either negligible or slightly negative.}
\label{tab:lora_scores_std}
\end{table}

\section{Results}

We are testing the impact of a large number of variables on a large number of metrics. In order to avoid p-hacking our way into incorrect conclusions, we take an unusually high threshold $p\leq0.001$ before we consider a result to be statistically significant.

\subsection{Social Media Fine-tuning Ablation Study}
\label{blueprint_results}

In order to isolate the impact of fine-tuning on the BluePrint dataset, we perform an ablation study using the Qwen2.5-7B-Instruct model. We run simulations with LoRA fine-tunings of this model on BluePrint (151 rollouts, randomized sim parameters), and contrast them with simulations run using only the model out-of-the-box, with a short prompt (\ref{appendix_prompt}) and a meta-personality randomly sampled from \citet{li2025llmgeneratedpersonapromise}. All other simulation parameters are randomized between runs. We chose Qwen specifically to perform this ablation study because it is the only instruction-tuned model among our selection, making it much easier to work with than the other base models. 

\begin{table}[h]
\centering
\begin{tabular}{lccc}
\toprule
Model & ($\downarrow$) Mean Accuracy $\pm \sigma$ & $n$ & 95\% C.I. \\
\midrule
Qwen2.5-7B-Instruct & $0.9999 \pm 0.0008$ & 72 & $[0.9935, 1.0000]$ \\
Qwen2.5-7B-Instruct-BluePrint (ours) & $\textbf{0.9531} \pm 0.0351$ & 151 & $[0.8487, 1.0000]$ \\
\midrule
 Llama-3.1-Minitaur-8B-BluePrint (ours) & $0.9544 \pm 0.0384$ & 169 & $[0.8387, 1.0000]$ \\
Llama-3.1-8B-BluePrint (ours) & $\textbf{0.9465} \pm 0.0475$ & 111 & $[0.7785, 1.0000]$ \\
 gemma-3-4b-pt-BluePrint (ours) & $0.9950 \pm 0.0083$ & 92 & $[0.9536, 1.0000]$ \\
\bottomrule
\end{tabular}
\caption{BERT human-vs-LLM classifier accuracy by model (lower is better). We observe that fine-tuning on BluePrint allows Qwen to be identified as AI less often. We also show the classifier's accuracy on our other models.}
\label{tab:bert_accuracy_by_model}
\end{table}

We first observe that the threads produced from non-finetuned models are almost always identified by the BERT human-vs-LLM classifier (0.9999 ± 0.0008 acc), which is much more than any of the models we finetuned (Table \ref{tab:bert_accuracy_by_model}). Without social media fine-tuning, the model writes in a characteristic "AI" style that is trivially distinguishable from real social media text. In contrast, the LoRA-equipped Qwen achieves 0.953 ($d=1.62$, $p<0.001$). While seemingly a modest improvement, it is important to contextualize this result. \cite{bückkaeffer2025textttblueprintsocialmediauser} showed that similar fine-tuning made humans judges nearly incapable of distinguishing humans from AI, and \cite{pagan2025computationalturingtestreveals} argue that BERT models are much better than humans at identifying the artificial nature of a text. Among LoRA-equipped models, Gemma-generated threads are classified with near-ceiling accuracy across all 92 runs, with minimal variance. The other three LoRA models produce text that is harder to distinguish (BERT accuracy ~95\%, with substantial run-to-run variance). Overall, we get Gemma vs every other BluePrint model: $d = 1.30–1.49$, all $p < 0.001$. The overall BERT accuracy is $\mu = 96.4\%$, with range [77.9\%, 100\%].

\begin{table}[h]
\centering
\begin{tabular}{lcccc}
\toprule
Metric 
& \makecell{Qwen2.5-7B-Instruct-\\BluePrint $\pm \sigma$} 
& \makecell{Qwen2.5-7B-\\Instruct $\pm \sigma$} 
& Cohen's $d$ 
& $p$ \\
\midrule
Opinion Shift Rate & $0.210 \pm 0.108$ & $0.057 \pm 0.097$ & 1.46 & $< 0.001$ *** \\
Majority Follow Rate & $0.505 \pm 0.064$ & $0.275 \pm 0.236$ & 1.60 & $< 0.001$ *** \\
NASR & $0.078 \pm 0.039$ & $0.021 \pm 0.036$ & 1.52 & $< 0.001$ *** \\
Net Consensus Change & $-0.055 \pm 0.161$ & $0.004 \pm 0.048$ & 0.44 & $< 0.001$ *** \\
\bottomrule
\end{tabular}
\caption{Simulation metrics with vs. without BluePrint fine-tuning. Finetuned models display more intense opinion dynamics.}
\label{tab:qwen_lora_comparison}
\end{table}

Without persona LoRAs, agents change opinions at one quarter the rate (OSR 0.057 vs 0.210), follow the majority half as often (MFR 0.275 vs 0.505), and produce essentially zero net change in group consensus (Table \ref{tab:qwen_lora_comparison}). The simulation runs, but the social dynamics appear heavily dampened. Without opinion change, network structure is preserved rather than mixed. Qwen2.5-7B-Instruct has dramatically higher local agreement (0.960 vs 0.722, d = 1.74) and lower cross-cutting edge fraction (0.040 vs 0.282, d = 1.74). This is not evidence of echo chamber formation, it reflects both the agents' unanimity and the absence of the opinion dynamics that would otherwise disrupt it. 
Qwen2.5-7B-Instruct starts every simulation at near-perfect consensus (init = 1.000 for Q25 and Q28, 0.907 for Q29).
The base model, without persona conditioning, produces the same "default" opinion on each topic from every agent despite the meta-personas. The persona LoRAs are what introduce the heterogeneity of starting positions that makes disagreement, persuasion, and consensus change happen.

It is important to keep in mind that without ground-truth comparison, we may not affirm that one of these approaches is more realistic than the other; we may only perform relative comparisons between the various possible setups. Additionally, these results go against much of the existing literature in which a large number of similar simulators have obtained convincing opinion dynamics without fine-tuning (e.g., \cite{yang2025oasisopenagentsocial}). This discrepancy is likely due to the shorter time horizons in our simulations. However, given that BluePrint produces more believable threads and more accentuated opinion dynamics, we perform the rest of our analysis using these finetuned models.

\subsection{Varying Base Models}

\begin{table}[h]
\centering
\small
\setlength{\tabcolsep}{4pt}

\begin{tabularx}{\linewidth}{lcccc}
\toprule
\makecell[l]{Model (LoRAs)} 
& $n$ 
& \makecell{Initial Consensus} 
& \makecell{$\Delta$ Consensus $\pm\sigma$} 
& \makecell{Runs with $\downarrow$ Cons.} \\
\midrule
Llama-3.1-Minitaur-8B-BluePrint & 169 & 0.86 & $\mathbf{-0.095 \pm 0.112}$ & 83\% \\
Qwen2.5-7B-Instruct-BluePrint   & 151 & 0.85 & $\mathbf{-0.055 \pm 0.161}$ & 70\% \\
Llama-3.1-8B-BluePrint          & 111 & 0.79 & $\mathbf{-0.027 \pm 0.179}$ & 50\% \\
gemma-3-4b-pt-BluePrint         & 92  & 0.73 & $\mathbf{+0.025 \pm 0.122}$ & 36\% \\
\bottomrule
\end{tabularx}

\caption{Consensus change by model. Negative $\Delta$ indicates a decrease in consensus from one survey to the next. The model choice has a major impact on both the initial opinion distribution and on the way in which it evolves over time.}
\label{tab:consensus_shift}
\end{table}

All four models produce different consensus trajectories at different rates (table \ref{tab:consensus_shift}). Minitaur produces the strongest, most consistent consensus erosion; Gemma is the only model that drifts slightly toward consensus gain on average (though the difference from zero is not significant given its variance). The ranking (Minitaur $>$ Qwen $>$ Llama $>$ Gemma in consensus erosion) does not obviously correspond to model size or architecture family. Both Minitaur and Llama share the same Llama-3.1-8B base but differ substantially in behavior, suggesting fine-tuning on Minitaur's social media persona data \citep{binz2025foundation} instills distinctly stronger individual opinionation.

\subsection{Seeing Survey Answers}

If we let models know about their own answers to survey questions by adding it to their context, we note that BERT AI detectability becomes substantially higher: ctx=True: 0.984 ± 0.023 vs ctx=False: 0.943 ± 0.043 ($t = 14.2$, $p < 10^{-15}$, $d = 1.20$). This is easily explained by the fact that models tend to hyper-focus on the survey question if they know about it, to the point where every message ends up being about the survey's topic. It then becomes trivial to identify an AI thread.

Majority follow rate (fraction of opinion-changers adopting the majority view) is higher when agents see their own answers to survey questions in their context: ctx=True: $0.495 \pm 0.111$ vs ctx=False: $0.461 \pm 0.138$ ($t = 3.30$, $p = 0.001$, Cohen's $d = 0.27$). However, these effects are not uniform across models. Table \ref{tab:context_effects} breaks down the survey context effect on consensus change within each model.

\begin{table}[h]
\centering
\begin{tabular}{lccc}
\toprule
Model & ctx=True & ctx=False & $p$ \\
\midrule
Llama-3.1-Minitaur-8B-BluePrint & $-0.095$ & $-0.095$ & $0.99$ (ns) \\
Llama-3.1-8B-BluePrint & $+0.015$ & $-0.067$ & $0.014$ * \\
Qwen2.5-7B-Instruct-BluePrint & $-0.004$ & $-0.107$ & $< 0.001$ *** \\
gemma-3-4b-pt-BluePrint & $+0.042$ & $+0.002$ & $0.11$ (ns) \\
\bottomrule
\end{tabular}
\caption{Effect of adding survey context on consensus change by model. The model and its context interact in seemingly unpredictable ways: letting models know about their survey answers could significantly change the results, depending on the model.}
\label{tab:context_effects}
\end{table}

Survey context has no effect on Minitaur or Gemma consensus, but significantly reduces consensus erosion for Qwen (p $<$ 0.001). When Qwen agents can see their own votes, they conform more and the aggregate consensus stabilizes or drifts upward. Minitaur agents apparently do not substantially adjust their behavior based on this information. Similarly, the BERT detectability uplift from survey context is almost entirely absent for Gemma (99.8\% vs 99.1\%), and is driven by the non-Gemma models (98.1\% vs 93.6\%).

\subsection{Geometry of the Parameter Space}

\label{geometry_parameter_space}

We investigate the interactions between variables to determine if the \textit{Silicon Society} design space behaves additively (knowing A increases X and B increases X implies A+B increases X more than either alone). Our findings suggest a hybrid geometry where certain parameters act independently, while others exhibit non-linear interactions gated by model identity or population scale. Note that we use the word "geometry" loosely: we do not formally define a space, we simply perform an empirical characterization of parameter interactions.

\textbf{Additive Effects.} Several parameters display consistent impacts across different simulation settings. For instance, the effect of homophily on assortativity shift is statistically significant and maintains a consistent direction regardless of the base model or population size. Similarly, the impacts of providing survey context and the presence of a news agent on BERT detectability are independent (interaction contrast IC=-0.006), indicating that they contribute additively to stylistic realism.

\textbf{Non-Additive and Synergistic Interactions.} In contrast, some parameters exhibit dependencies that preclude simple extrapolation from marginal effects. The influence of survey context on consensus change is highly model-dependent: it serves as a robust driver for \textit{Qwen} and \textit{Llama} but has no measurable effect on \textit{Minitaur} or \textit{Gemma} (Table \ref{tab:context_effects}). Furthermore, the effect of survey context on the Majority Follow Rate (MFR) is synergistic with population size; while negligible at small scales ($N\leq256$), the effect becomes substantial and robust as the population reaches 1024 agents or more ($p<0.001$, Table \ref{tab:ctx_effect_mfr}).

\begin{table}[h]
\centering
\begin{tabular}{lcc}
\toprule
num\_agents & Effect of survey context on MFR & $p$ \\
\midrule
64    & $-0.020$ & 0.36 (ns) \\
256   & $+0.024$ & 0.25 (ns) \\
1024  & $+0.075$ & $< 0.001$ *** \\
4096  & $+0.064$ & 0.001 *** \\
\bottomrule
\end{tabular}
\caption{Effect of adding survey context on majority follow rate (MFR) across different numbers of agents.}
\label{tab:ctx_effect_mfr}
\end{table}

\subsection{Analysis of Variance}

In order to identify which parameters matter most when designing an LLM-based social simulator, we compute which parameters explain most of the variance for each metric (table \ref{tab:anova_factors}). The sum of all individual $\eta^2$ values falls well below 1.0 for each metric, reflecting both residual noise and potential interactions between parameters. We observe that the number of agents seems to have an important impact on Opinion Shirt Rate, however it is likely an artifact of the way in which we survey models (see \ref{appendix_osr_vs_num_agents}). On the other hand, presence or absence of a biased news agent has no significant effect on consensus change ($p = 0.62$), opinion shift rate ($p = 0.23$), or BERT detectability ($p = 0.34$). A single news agent does not measurably alter simulation dynamics. This could be because the signal from a single agent is too weak to have an impact, or because network topology is limiting spread.

\begin{table}[h]
\centering
\begin{tabular}{lccc}
\toprule
Metric & Dominant Factor & $\eta^2$ & Secondary Factor \\
\midrule
Opinion Shift Rate & num\_agents & $\mathbf{0.230}$ & model (0.200) \\
Net Consensus Change & model & $\mathbf{0.090}$ & ctx (0.028) \\
BERT accuracy & model & $\mathbf{0.266}$ & ctx (0.264) \\
$\Delta$ assortativity & homophily & $\mathbf{0.088}$ & num\_agents(0.037) \\
\bottomrule
\end{tabular}
\caption{Variance explained by each parameter individually ($\eta^2$, single-factor)}
\label{tab:anova_factors}
\end{table}

\section{Discussion}

\subsection{Conclusion}

We conducted a systematic exploration of the design space for LLM-based social simulations, which we refer to as \textit{Silicon Societies}. By conducting 595 simulation roll-outs across an array of parameters (e.g. base models, population proportions, network topologies) we identified several key factors that define the behavior of these simulated environments.

We found that the geometry of the design space is neither trivially additive nor uniformly complex. The interactions between parameters, such as the relationship between homophily and model-specific tendencies, suggest that researchers cannot treat simulation settings as independent variables. Most notably, we found that the choice of the base model is the single most influential factor in determining simulation outcomes, outweighing structural factors like network topology. Furthermore, our experiments with the BluePrint dataset demonstrate that fine-tuning models on real-world social media data constitutes a step towards achieving stylistic realism and stronger opinion dynamics.

As the field moves toward more complex agentic technologies, this study offers a baseline for more informed design decisions. Future work should focus on expanding this design space to include more nuanced action spaces and exploring the longitudinal stability of these digital societies. Ultimately, understanding these design constraints is a prerequisite for closing the validation gap and transforming \textit{Silicon Societies} into reliable tools for social science research. We hope this analysis provides a foundation upon which future works can build more principled simulators, and encourages the community to adopt more rigorous, multi-parameter evaluation frameworks. 

\subsection{Limitations}

The parameter space is immense: not only are there many possible parameters we did not cover, even in the parameters we covered we did not cover nearly all possible options. We hope our approach will inspire similar works to explore more of the parameter space.

The base model variable is confounded with question-specific initial consensus values (e.g., Qwen on Q28 starts at 0.993, Gemma on Q29 at 0.562), which may affect within-question comparisons.

Statistical tests are unadjusted for multiple comparisons. With ~30+ tests performed, several nominally significant results at $p < 0.05$ may be false positives; treat marginal results ($p < 0.05$) with caution. Once again, we reiterate that we only consider $p < 0.001$ to be statistically significant in this study, and advise readers do the same. p-values are computed using Welch's t test. We use a normal approximation rather than a t-distribution CDF. This is accurate for moderate-to-large samples but slightly less accurate for small subgroups.

We do not consider effects resulting from the interaction of more than two parameters because the number of possibilities to consider explodes quickly.

\section*{Acknowledgments}

Some of the data analysis in this paper was performed using Claude code (Sonnet 4.6).

\bibliography{colm2026_conference}
\bibliographystyle{colm2026_conference}

\appendix

\section{Appendix}

\subsection{Simulator details}
\label{appendix_simulator_details}

This social media simulator is designed for the study of collective opinion dynamics and information diffusion within synthetic social networks. The system integrates LLMs architectures with established network science principles to model the emergence of polarization, consensus, and information cascades.

The simulator operates on a discrete-time execution model managed by a central engine (SimEngine). Each simulation step encompasses a multi-stage observation and action cycle (10 observations followed by a single action), ensuring that agents are sufficiently exposed to social context before generating content. To mirror empirical social media engagement patterns, agent activation follows a Zipfian distribution, where a minority of highly active "power users" drive the majority of network interactions.

Agents are implemented as autonomous entities powered by LLMs, utilizing the Unsloth framework for optimized local inference. Behavioral heterogeneity is achieved through a multi-adapter architecture, where a shared base model (e.g., Llama-3) is augmented with diverse Low-Rank Adaptation (LoRA) modules. These adapters allow for the efficient simulation of distinct personality profiles and cognitive biases across the agent population. Furthermore, agents can be assigned specific personas that inform their system prompts, enabling research into how individualized traits influence collective behavior.

The environment supports flexible social graph configurations, primarily focusing on Power-law Cluster Graphs. This topology captures the essential features of real-world social networks, including scale-free degree distributions and high local clustering. The framework allows for the controlled study of structural homophily, where agents are positioned within the network based on the alignment of their underlying belief systems. External information is introduced via "News Source" nodes, which inject curated datasets (e.g., political or environmental tweet corpuses) into high-degree positions in the graph to simulate the impact of media exposure on the network.

The simulator features a comprehensive suite of metrics to quantify social phenomena:

Echo Chamber Dynamics: Measurements of assortativity, local agreement, and the prevalence of cross-cutting edges that bridge opposing ideological clusters.
Opinion Evolution: Longitudinal tracking of agent responses to standardized survey questions, providing a ground-truth trajectory of belief shifts.
Information Diffusion: Analysis of "herd effects," neighbor alignment, and the evolution of consensus or fragmentation over time.

Agents are surveyed every 250 steps, starting at step 0 and ending at step 2500 for a total of 11 surveys per simulation. When surveying models, we take the answer with the largest log-probability as being a given agent's choice in order to reduce noise.

\subsection{Simulation counts for every setting}
\label{appendix_sim_counts}

\begin{table}[h]
\centering
\begin{tabular}{llcc}
\toprule
Category & Value & $n$ & Total \\
\midrule
\multirow{5}{*}{Model}
 & Llama-3.1-Minitaur-8B-BluePrint & 169 & \multirow{5}{*}{595} \\
 & Llama-3.1-8B-BluePrint & 111 & \\
 & Qwen2.5-7B-Instruct-BluePrint & 151 & \\
 & Qwen2.5-7B-Instruct & 72 & \\
 &  gemma-3-4b-pt-BluePrint & 92 & \\
\midrule
\multirow{4}{*}{Num Agents}
 & 64 & 155 & \multirow{4}{*}{595} \\
 & 256 & 163 & \\
 & 1024 & 144 & \\
 & 4096 & 133 & \\
\midrule
\multirow{3}{*}{Question}
 & Q25 & 204 & \multirow{3}{*}{595} \\
 & Q28 & 192 & \\
 & Q29 & 199 & \\
\midrule
\multirow{2}{*}{Graph Type}
 & powerlaw\_cluster & 308 & \multirow{2}{*}{595} \\
 & random & 287 & \\
\midrule
\multirow{2}{*}{Homophily}
 & False & 312 & \multirow{2}{*}{595} \\
 & True & 283 & \\
\midrule
\multirow{2}{*}{Survey in Context}
 & False & 283 & \multirow{2}{*}{595} \\
 & True & 312 & \\
\midrule
\multirow{5}{*}{Proportions Option}
 & None (Qwen2.5-7B-Instruct) & 72 & \multirow{5}{*}{595} \\
 & average & 118 & \\
 & blueprint & 152 & \\
 & distribution & 115 & \\
 & uniform & 138 & \\
\midrule
\multirow{2}{*}{Num News Agents}
 & 0 & 312 & \multirow{2}{*}{595} \\
 & 1 & 283 & \\
\bottomrule
\end{tabular}
\caption{Summary of simulation counts across all experimental dimensions. In total, we ran 595 simulations.}
\label{tab:all_counts_total}
\end{table}

\subsection{Qwen2.5-7B-Instruct (without BluePrint lora fine-tuning) prompt}
\label{appendix_prompt}

"You are a user on a social media platform. Write a new post, or a comment in response to a thread. Only write in character. Speak in english, and answer in a style consistent with the following persona: " + persona

Example of a meta-persona from \cite{li2025llmgeneratedpersonapromise}: '\{"AGE": "45-54", "SEX": "Male", "RACE": "White", "STATE": "NJ"\}'

The agent's in-simulation name was selected to match the gender described by the persona.

\subsection{Opinion Shift Rate VS Number of Agents}
\label{appendix_osr_vs_num_agents}

While we note a very clear trend in which opinion shift rate decreases in relation to the number of agents in the sim (Appendix, Table \ref{tab:num_agents_results}), this is very likely an artifact of the way in which we survey models: we survey every 250 steps, but with more agents, a smaller fraction of them may have had a chance to act/observe in this time span, hence less opinion changes happen.

\begin{table}[h]
\centering
\begin{tabular}{lcccc}
\toprule
num\_agents 
& \makecell{Llama-3.1-Minitaur-\\8B-BluePrint}
& \makecell{Llama-3.1-8B-\\BluePrint}
& \makecell{Qwen2.5-7B-\\Instruct-BluePrint}
& \makecell{gemma-3-4b-\\pt-BluePrint} \\
\midrule
64   & 0.199 & 0.255 & 0.269 & 0.282 \\
256  & 0.230 & 0.267 & 0.274 & 0.315 \\
1024 & 0.110 & 0.151 & 0.160 & 0.178 \\
4096 & 0.060 & 0.129 & 0.149 & 0.117 \\
\bottomrule
\end{tabular}
\caption{Opinion shift rate per model per number of agents. This correlation is very likely an artifact of the frequency at which models are surveyed, and should be interpreted with caution.}
\label{tab:num_agents_results}
\end{table}

\subsection{Designing survey questions}
\label{appendix_designing_survey_questions}

We conjecture that if a survey question splits the studied population evenly, it is more likely to lead to clearly identifiable opinion dynamics than if there is an overwhelming majority of votes in favor of one side. Given our stated goal, which is to identify and quantify the impact of several design decisions on simulation outcomes, we pick questions that appear more likely to make any differences between roll-outs clearer even if there is an argument to be made on the interesting nature of dynamics involving a minority and a majority opinion.

In order to select questions that evenly split the population, we created a set of 42 potentially divisive multiple-choice questions using a mix of human and Gemini 3.1 suggestions (see \ref{appendix_survey_question_entropy}). We then take our population of finetuned loras $L$ and for each question $q$ compute the vector $\vec{v}$ defined such that for each possible answer $a_i$ to question $q$ we have

$$
v_i:=\frac{1}{|L|}\sum_{\ell\in L}w_{\ell}P(\ell(q)=a_i), \quad \mathcal{H}(\vec{v}):=-\sum_{i}v_i\operatorname{log_2}v_i
$$

Where $\mathcal{H}$ denotes Shannon entropy \citep{bromiley2004shannon}. Essentially we are taking a weighted average of population opinion, and by picking the questions that maximizes $\mathcal{H}$ we select for divisiveness (or indecisiveness, depending on the exact underlying distribution of agent opinions). While our approach allows questions with an arbitrary number of options, we only use binary choices for the sake of simplicity.

We find that the entropy of a given question depends heavily on both the base model on which the LoRAs are trained, as well as on the proportions of these LoRAs in the population. However, comparing the results of various simulation roll-outs involving different base models and model proportions is impossible if we also change the question we are using to survey them. We therefore pick 2 questions that showed the highest entropy for the \textit{Llama-3.1-Minitaur-8B-BluePrint} LoRAs, and the one that showed the lowest entropy (Table \ref{tab:selected_survey_questions}). We observe that while the population is very close to a 50-50 split on these questions, some individual models do show strong preferences towards either side instead of all being moderate.

\begin{longtable}{cp{7cm}ccc}
\caption{Selected survey questions, ranked by entropy} \label{tab:selected_survey_questions} \\
\toprule
\textbf{\#} & \textbf{Question} & \textbf{YOpt 1 (\%)} & \textbf{No / Opt 2 (\%)} & \textbf{Entropy} \\ \midrule
\endfirsthead

\multicolumn{5}{c}
{{\bfseries \tablename\ \thetable{} -- continued from previous page}} \\
\toprule
\textbf{\#} & \textbf{Question} & \textbf{Yes / Opt 1 (\%)} & \textbf{No / Opt 2 (\%)} & \textbf{Entropy} \\ \midrule
\endhead

\midrule
\multicolumn{5}{r}{{Continued on next page}} \\
\bottomrule
\endfoot

\bottomrule
\endlastfoot

1 & Should it be legal to use copyrighted material to train artificial intelligence models? You may only answer with 'Yes' or 'No'. & 48.46 & 51.54 & 0.9993 \\ \midrule
2 & Is it acceptable to use genetic engineering to enhance human intelligence? You may only answer with 'Yes' or 'No'. & 48.42 & 51.58 & 0.9993 \\ \midrule
42 & Would you rather vote for a candidate who prioritizes economic growth over environmental protection, or a candidate who prioritizes environmental protection over economic growth? You may only answer with 'Economic Growth' or 'Environmental Protection'. & 88.97 & 11.03 & 0.5008 \
\end{longtable}

We observe that simply changing the order in which options are presented in a question can have a dramatic impact on the distribution of answers (and therefore on the entropy). For example, "Would you rather vote for Donald Trump or
Kamala Harris? You may only answer with ’Donald Trump’ or ’Kamala Harris’." ranks 3rd in terms of entropy, while "Would you rather vote for Kamala Harris or Donald Trump? You may only answer with ’Kamala Harris’ or ’Donald Trump’." ranks 40th. This echoes known results in both humans and LLMs \citep{schuman1996questions,tjuatja2024llms}. 

\subsection{Survey questions entropy}
\label{appendix_survey_question_entropy}

\begin{longtable}{cp{7cm}ccc}
\caption{Survey questions, ranked by entropy.} \label{tab:survey_data} \\
\toprule
\textbf{\#} & \textbf{Question} & \textbf{Yes / Opt 1 (\%)} & \textbf{No / Opt 2 (\%)} & \textbf{Entropy} \\ \midrule
\endfirsthead

\multicolumn{5}{c}%
{{\bfseries \tablename\ \thetable{} -- continued from previous page}} \\
\toprule
\textbf{\#} & \textbf{Question} & \textbf{Yes / Opt 1 (\%)} & \textbf{No / Opt 2 (\%)} & \textbf{Entropy} \\ \midrule
\endhead

\midrule
\multicolumn{5}{r}{{Continued on next page}} \\
\bottomrule
\endfoot

\bottomrule
\endlastfoot

1 & Should it be legal to use copyrighted material to train artificial intelligence models? You may only answer with 'Yes' or 'No'. & 48.46 & 51.54 & 0.9993 \\ \midrule
2 & Is it acceptable to use genetic engineering to enhance human intelligence? You may only answer with 'Yes' or 'No'. & 48.42 & 51.58 & 0.9993 \\ \midrule
3 & Would you rather vote for Donald Trump or Kamala Harris? You may only answer with 'Donald Trump' or 'Kamala Harris'. & 48.22 & 51.78 & 0.9991 \\ \midrule
4 & Should the government provide free college education for all citizens? You may only answer with 'Yes' or 'No'. & 47.31 & 52.69 & 0.9979 \\ \midrule
5 & Should the government implement a universal basic income? You may only answer with 'Yes' or 'No'. & 52.87 & 47.13 & 0.9976 \\ \midrule
6 & Is it acceptable to use performance-enhancing drugs in professional sports? You may only answer with 'Yes' or 'No'. & 47.09 & 52.91 & 0.9976 \\ \midrule
7 & Is it ethical to use artificial intelligence in military applications? You may only answer with 'Yes' or 'No'. & 54.10 & 45.90 & 0.9951 \\ \midrule
8 & Should the voting age be lowered to 16? You may only answer with 'Yes' or 'No'. & 54.57 & 45.43 & 0.9940 \\ \midrule
9 & Should the death penalty be abolished worldwide? You may only answer with 'Yes' or 'No'. & 54.80 & 45.20 & 0.9934 \\ \midrule
10 & Should the government implement stricter gun control laws? You may only answer with 'Yes' or 'No'. & 55.16 & 44.84 & 0.9923 \\ \midrule
11 & Is it acceptable to genetically modify human embryos? You may only answer with 'Yes' or 'No'. & 44.41 & 55.59 & 0.9910 \\ \midrule
12 & Is it ethical to use human subjects for medical research without their informed consent? You may only answer with 'Yes' or 'No'. & 44.33 & 55.67 & 0.9907 \\ \midrule
13 & Would you rather vote for Pierre Poilievre or Mark Carney? You may only answer with 'Pierre Poilievre' or 'Mark Carney'. & 43.91 & 56.09 & 0.9893 \\ \midrule
14 & Is it acceptable to censor hate speech on social media platforms? You may only answer with 'Yes' or 'No'. & 56.20 & 43.80 & 0.9889 \\ \midrule
15 & Would you rather vote for Donald Trump or Joe Biden? You may only answer with 'Donald Trump' or 'Joe Biden'. & 56.28 & 43.72 & 0.9886 \\ \midrule
16 & Is it ethical to use animals for food production? You may only answer with 'Yes' or 'No'. & 56.48 & 43.52 & 0.9879 \\ \midrule
17 & Is it ethical to use human cloning for reproductive purposes? You may only answer with 'Yes' or 'No'. & 42.16 & 57.84 & 0.9822 \\ \midrule
18 & Should the government implement a carbon tax to combat climate change? You may only answer with 'Yes' or 'No'. & 58.35 & 41.65 & 0.9798 \\ \midrule
19 & Is it acceptable to use surveillance technology to monitor public spaces for security purposes? You may only answer with 'Yes' or 'No'. & 41.65 & 58.35 & 0.9798 \\ \midrule
20 & Should the government implement a universal basic income for artists and creatives? You may only answer with 'Yes' or 'No'. & 58.44 & 41.56 & 0.9793 \\ \midrule
21 & Should the government implement a wealth tax on the richest individuals? You may only answer with 'Yes' or 'No'. & 59.08 & 40.92 & 0.9761 \\ \midrule
22 & Is it acceptable to create separate categories for athletes using performance-enhancing drugs in professional sports? You may only answer with 'Yes' or 'No'. & 40.37 & 59.63 & 0.9731 \\ \midrule
23 & Should the government provide free childcare for all citizens? You may only answer with 'Yes' or 'No'. & 59.88 & 40.12 & 0.9717 \\ \midrule
24 & Is it ethical to use animals for scientific research? You may only answer with 'Yes' or 'No'. & 60.48 & 39.52 & 0.9681 \\ \midrule
25 & Is it ethical to use embryonic stem cells for medical research? You may only answer with 'Yes' or 'No'. & 60.87 & 39.13 & 0.9656 \\ \midrule
26 & Would you rather vote for a candidate who supports private healthcare or a candidate who supports universal healthcare? You may only answer with 'Private Healthcare' or 'Universal Healthcare'. & 36.39 & 63.61 & 0.9459 \\ \midrule
27 & Would you rather vote for a candidate who supports looser gun control laws or a candidate who supports stricter gun control laws? You may only answer with 'Looser Gun Control Laws' or 'Stricter Gun Control Laws'. & 33.55 & 66.45 & 0.9204 \\ \midrule
28 & Should the government provide free healthcare for all citizens? You may only answer with 'Yes' or 'No'. & 66.77 & 33.23 & 0.9173 \\ \midrule
29 & Should recreational marijuana be legalized? You may only answer with 'Yes' or 'No'. & 67.04 & 32.96 & 0.9145 \\ \midrule
30 & Is it acceptable to use facial recognition technology in public spaces? You may only answer with 'Yes' or 'No'. & 32.86 & 67.14 & 0.9135 \\ \midrule
31 & Would you rather vote for Bob or John? You may only answer with 'Bob' or 'John'. & 31.35 & 68.65 & 0.8972 \\ \midrule
32 & Is it ethical to use animals for entertainment purposes, such as in circuses or zoos? You may only answer with 'Yes' or 'No'. & 29.97 & 70.03 & 0.8809 \\ \midrule
33 & Should the government provide free mental health services for all citizens? You may only answer with 'Yes' or 'No'. & 70.60 & 29.40 & 0.8738 \\ \midrule
34 & Would you rather vote for a candidate who prioritizes environmental protection over economic growth, or a candidate who prioritizes economic growth over environmental protection? You may only answer with 'Environmental Protection' or 'Economic Growth'. & 29.07 & 70.93 & 0.8696 \\ \midrule
35 & Would you rather vote for John or Bob? You may only answer with 'John' or 'Bob'. & 71.73 & 28.27 & 0.8591 \\ \midrule
36 & Would you rather vote for a candidate who supports stricter gun control laws or a candidate who supports looser gun control laws? You may only answer with 'Stricter Gun Control Laws' or 'Looser Gun Control Laws'. & 80.25 & 19.75 & 0.7169 \\ \midrule
37 & Would you rather vote for a candidate who supports universal healthcare or a candidate who supports private healthcare? You may only answer with 'Universal Healthcare' or 'Private Healthcare'. & 82.21 & 17.79 & 0.6755 \\ \midrule
38 & Would you rather vote for Joe Biden or Donald Trump? You may only answer with 'Joe Biden' or 'Donald Trump'. & 82.21 & 17.79 & 0.6755 \\ \midrule
39 & Is climate change primarily caused by human activity? You may only answer with 'Yes' or 'No'. & 83.34 & 16.66 & 0.6498 \\ \midrule
40 & Would you rather vote for Kamala Harris or Donald Trump? You may only answer with 'Kamala Harris' or 'Donald Trump'. & 84.56 & 15.44 & 0.6207 \\ \midrule
41 & Would you rather vote for Mark Carney or Pierre Poilievre? You may only answer with 'Mark Carney' or 'Pierre Poilievre'. & 88.57 & 11.43 & 0.5128 \\ \midrule
42 & Would you rather vote for a candidate who prioritizes economic growth over environmental protection, or a candidate who prioritizes environmental protection over economic growth? You may only answer with 'Economic Growth' or 'Environmental Protection'. & 88.97 & 11.03 & 0.5008 \
\end{longtable}

\subsection{Homophily}
\label{appendix_homophily}

In this simulator, graph initialization with homophily is implemented as a spatial-opinion alignment strategy. The process ensures that agents with similar ideological profiles are positioned in the same "neighborhoods" of the social graph.

The exact algorithm follows these steps:

\begin{enumerate}
    \item \textbf{Spatial Embedding}: We first computes a 2D spatial layout of the social graph (Power-law Cluster Graph) using the Fruchterman-Reingold spring algorithm (`nx.spring\_layout`). This algorithm naturally clusters well-connected nodes together in Euclidean space.
    \item \textbf{Node Ordering}: All nodes in the graph are then sorted based on their X-coordinate from this spatial layout. This creates a one-dimensional "left-to-right" ordering of the network's structural regions.
    \item \textbf{Opinion Scoring}: For every agent (or LLM adapter profile), the system calculates an expected opinion score based on its baseline probability distribution for specific survey questions (e.g., scoring a "Strongly Disagree" as 0 and "Strongly Agree" as 4).
    \item \textbf{Agent Sorting}: The agent population is grouped by their model profiles and sorted by their mean opinion scores, creating an ideological spectrum from one extreme to the other.
    \item \textbf{Interleaved Assignment}: Finally, the spatially ordered nodes are mapped directly to the opinion-sorted agents. 
\end{enumerate}

\textbf{Result}: Agents with the lowest opinion scores are assigned to the "leftmost" nodes of the graph's spatial layout, while agents with the highest scores are assigned to the "rightmost" nodes. Because the spring layout places connected nodes near each other, this mapping ensures that neighbors in the social graph are significantly more likely to share similar initial opinions, creating the initial structural homophily required for studying echo chamber formation.

\subsection{Metrics}
\label{appendix_metrics}

\begin{table}[h]
\centering
\small
\begin{tabular}{|l|c|p{5cm}|c|}
\hline
\textbf{Metric} & \textbf{Acronym} & \textbf{Explanation} & \textbf{Formula} \\ \hline
Consensus & --- & The degree of global agreement; the proportion of the population holding the most popular opinion at a specific time step. & $\frac{N_{\max}}{N_{\text{total}}}$ \\ \hline
Net Consensus Change & NCC & The total convergence or divergence of the population over the entire simulation duration ($t_{final} - t_{0}$). & $C_{t_f} - C_{t_0}$ \\ \hline
Opinion Shift Rate & OSR & The fraction of the population that changed their opinion between two consecutive survey snapshots. & $\frac{N_{\text{changed}}}{N_{\text{total}}}$ \\ \hline
Majority Follow Rate & MFR & Of the agents who changed their minds, the fraction that adopted the opinion that is currently the majority. & $\frac{N_{\text{changed} \to \text{maj}}}{N_{\text{changed}}}$ \\ \hline
Neighbor Alignment Shift Rate & NASR & The fraction of the total population that changed their opinion specifically to match the majority of their direct social neighbors. & $\frac{N_{\text{changed} \to \text{nb\_maj}}}{N_{\text{total}}}$ \\ \hline
Assortativity & --- & A measure of homophily; strictly, the tendency of agents to be connected to others with the same opinion (Network Assortativity Coefficient). & $\frac{\text{Tr}(\mathbf{e}) - \lVert \mathbf{e}^2 \rVert}{1 - \lVert \mathbf{e}^2 \rVert}$ \\ \hline
\end{tabular}
\caption{Behavioral and Network Metrics defined in the Social-Sim Engine.}
\label{tab:metrics_definitions}
\end{table}

\subsection{Software Used}

Our experiments relied on Transformers \citep{wolf-etal-2020-transformers}, NumPy \citep{harris2020array}, PyTorch \citep{Ansel_PyTorch_2_Faster_2024}, NetworkX \citep{hagberg2008networkx} and Unsloth \citep{unsloth}.

\end{document}